\documentclass[12pt]{iopart}

\usepackage{graphicx}
\usepackage{dcolumn}
\usepackage{bm}
\usepackage{subfigure}
\usepackage[utf8]{inputenc}
\usepackage[T1]{fontenc}
\usepackage{epstopdf}
\expandafter\let\csname equation*\endcsname\relax
\expandafter\let\csname endequation*\endcsname\relax
\usepackage{amsthm,amsmath,amssymb,amsfonts}
\usepackage{mathrsfs}
\usepackage{ulem}
\usepackage{soul}
\usepackage{cite}
\usepackage{color}
\usepackage{soul}
\usepackage{float}
\usepackage{threeparttable}
\usepackage{booktabs}
\usepackage{multirow}
\usepackage{upgreek}
\usepackage[justification=centering]{caption}
\large
\usepackage{lineno}

\begin{document}

\title[Unified mesoscale picture of nonlinear generation of zonal flows in toroidal geometry]{Unified mesoscale picture of nonlinear generation of zonal flows in toroidal geometry}

\author{Zihao Wang and Shaojie Wang}
\address{Department of Engineering and Applied Physics, University of Science and Technology of China, Hefei, 230026, China}
\ead{wangsj@ustc.edu.cn}
\vspace{10pt}
\begin{indented}
\item[]July 2024
\end{indented}

\begin{abstract}
  Recent experimental findings on limit-cycle-oscillations indicate that the nonlinear driving of the turbulent poloidal Reynolds stress to zonal flows is not a significant factor in the toroidal geometry, sparking fundamental controversial issues within the fusion community.
  By using the global nonlinear gyrokinetic simulations, we propose a unified mesoscale picture of nonlinear driving of zonal flows in the ion-temperature-gradient turbulence.
  Zonal flows are nonlinearly driven by the turbulent energy flux and the turbulent poloidal Reynolds stress.
  The turbulent energy flux is not shielded by the toroidal geometry effect in nonlinearly driving zonal flows.
  The turbulent poloidal Reynolds stress is not shielded on the time scale shorter than the ion bounce period; however, on the time scale longer than the ion bounce period, the turbulent poloidal Reynolds stress is indeed shielded by the toroidal geometry effect.
\end{abstract}

%
%
%
\maketitle
%
%

\section{Introduction}

Mesoscale structures, commonly referred to as Zonal Flows (ZFs)~\cite{hasegawa1987self,diamond2005zonal}, are ubiquitous in various natural environments and geometries.
These encompass zonally symmetric structures in spherical geometry, such as extratropical circulation~\cite{thompson2000annular,l2006observed} in meteorology and atmospheric superrotation~\cite{schubert1983general} on Venus, as well as vortex structures in near-cylindrical geometry like hurricanes formed over vast oceans~\cite{guinn1993hurricane} and the famous Great Red Spot on Jupiter~\cite{marcus1988numerical,porco2003cassini}.
In magnetic confinement fusion plasmas, ZFs under toroidal geometry have been extensively observed~\cite{lin1998turbulent,diamond2005zonal,itoh2006physics,fujisawa2008review}, and are believed to be a crucial role in suppressing turbulence and improving confinement in tokamaks, e.g., the transition from Low- to High-confinement (L-H transition)~\cite{wagner1982regime} and the formation of internal transport barriers~\cite{connor2004review,ida2018internal}.
During the L-H transition phase at the plasma edge, the rapidly varying ion pressure gradient, zonal radial electric field, and poloidal flow were simultaneously observed; however establishing the causality between them posed an early challenge.
Furthermore, theories about the dynamical processes involved in the nonlinear excitation of ZFs and their intricate interactions with turbulence are also constantly developing~\cite{itoh1988model,shaing1989bifurcation,biglari1990influence,diamond1991theory,hahm1999shearing,chen2000excitation,kim2003zonal,chen2004zonal}.
The study of ZFs dynamics not only enhances our comprehension of fusion plasma confinement, but also provides an excellent framework for exploring highly complex nonlinear systems across multi-scale.

From the microscopic view, linearly stable ZFs can be spontaneously driven by the modulational instability~\cite{chen2000excitation,chen2012nonlinear}.
Recent studies in the nonlinear gyrokinetic theory and simulations have shown that ZFs can also be nonlinearly driven by the eigenmode self-interaction~\cite{wang2022nonlinear}.
Note that the growth rate of the ZF driven by the modulational instability depends on the amplitude of the pumping wave, which is the linear eigenmode, while the growth rate of the ZF driven by the eigenmode self-interaction is exactly twice as large as that of the eigenmode instability.
It has been shown that the self-interaction dominates the nonlinear driving of ZFs during the quasilinear stage of Ion-Temperature-Gradient (ITG) mode, while the modulational instability is important during the nonlinear saturation of ITG turbulence~\cite{wang2022nonlinear}.

From the mesoscopic view, ZFs are nonlinearly driven by the transport fluxes.
The nonlinear driving mechanisms of the zonal radial electric field ($\delta E_{r}$) can be argued from the ion radial force balance equation, which is written as
\begin{equation}
    \delta E_{r}=\frac{1}{n e}\partial_{r} \delta p_{i}+\delta u_{\zeta}B_{\text{P}}-\delta u_{\theta}B_{\text{T}}. \label{eq:radial_force_balance}
\end{equation}
Here, $\delta u_{\theta}$ and $\delta u_{\zeta}$ denote the poloidal and toroidal component of ion fluid velocity, respectively; $\delta p_i$ denotes the ion pressure; $B_{\text{P}}$ and $B_{\text{T}}$ denote the poloidal and toroidal magnetic field, respectively; $r$ is the minor radius; $n$ and $e$ are ion density and charge, respectively.
According to the fluid model, the momentum and energy equations reveal the relation between fluid moments (such as $\delta u_{\theta}$, $\delta u_{\zeta}$, and $\delta p_{i}$) and transport fluxes (such as the turbulent poloidal Reynolds Stress (RS) $\Pi_{r\theta}$, the turbulent toroidal RS $\Pi_{r\zeta}$, and the turbulent energy flux $Q_{r}$).
These can be written as
\begin{subequations} \label{eq:transport_equations}
    \begin{align}
        \partial_{t} \delta u_{\theta} + \frac{1}{nm} \frac{1}{r}\partial_{r}\left( r\Pi_{r\theta}\right) = 0, \label{eq:poloidal RS} \\
        \partial_{t} \delta u_{\zeta} + \frac{1}{nm} \frac{1}{r}\partial_{r}\left( r\Pi_{r\zeta}\right) = 0, \label{eq:toroidal RS} \\
        \frac{2}{3} \partial_{t} \delta p_{i} + \frac{1}{r} \partial_{r} \left( rQ_{r}\right) = 0. \label{eq:turbulent energy flux}
    \end{align}
\end{subequations}
Here, $\Pi_{r\theta}=nm\langle \tilde{V}_{r}\tilde{V}_{\theta} \rangle_{\text{en}} $ and $\Pi_{r\zeta}=nm\langle \tilde{V}_{r}\tilde{V}_{\zeta} \rangle_{\text{en}} $ denote the turbulent poloidal RS and toroidal RS, respectively, which are related to the radial ($\tilde{V}_{r}$), poloidal ($\tilde{V}_{\theta}$), and toroidal ($\tilde{V}_{\zeta}$) components of the fluctuating $\bm{E}\times\bm{B}$ velocity, with $\langle\cdot\rangle_{\text{en}}$ being the ensemble average operator; $m$ is the ion mass.
From the combination of equations~\eqref{eq:radial_force_balance} and  ~\eqref{eq:transport_equations}, the nonlinear driving equation for ZFs in the cylindrical geometry can be derived in a fluid framework:
\begin{equation} \label{eq:drive_without_sheild}
\begin{split}
    \partial_{t} \delta E_{r} =&~ \frac{B_{\text{T}}}{nm}\frac{1}{r}\partial_{r}\left( r\Pi_{r\theta} \right) - \frac{B_{\text{P}}}{nm}\frac{1}{r}\partial_{r}\left( r\Pi_{r\zeta} \right)  \\
    &-\frac{1}{ne}\partial_{r}\left[ \frac{1}{r}\partial_{r}\left( r\frac{2}{3}Q_{r} \right) \right]. 
\end{split}
\end{equation}
The first term on the right-hand-side explains the nonlinear generation of ZFs by the turbulent poloidal RS~\cite{diamond1991theory}; note that the last two terms were ignored in Ref.~\cite{diamond1991theory}.
This viewpoint is partially confirmed by experimental observations~\cite{kim1994rotation,xu2000role} and simulation results~\cite{carreras1993resistive}, especially on the correlation between the change of $\delta E_{r}$ and the turbulent poloidal RS term.

The Limit-Cycle-Oscillations (LCO) near the L-H transition threshold allow experimentalists to measure the evolution of zonal radial electric field on longer time scale.
Many LCO experiments confirm the poloidal RS model, particularly regarding the correlation between the zonal radial electric field and the turbulent poloidal RS~\cite{conway2011mean,xu2011first,xu2012frequency,schmitz2012role}.
However, recent controversial reports in JFT-2M~\cite{kobayashi2013spatiotemporal,kobayashi2014dynamics} and HL-2A~\cite{cheng2013dynamics,cheng2014low} tokamaks argue that the turbulent poloidal RS is not important in nonlinear driving ZFs in the toroidal geometry due to the neoclassical shielding factor.
On the other hand, the ion pressure gradient is identified as significantly changing during the LCO phase in almost all tokamak devices.
Experimental observations of LCO characteristics have revealed the existence of multiple LCO states, where the I-phase is thought to be dominated by the ion pressure gradient, but analysis of mesoscale dynamics is still lacking~\cite{grover2023experimentally}.
In Refs.~\cite{kobayashi2013spatiotemporal,kobayashi2014dynamics}, it is argued that
\begin{equation}
    \partial_{t} \delta u_{\theta} + \frac{1}{\varepsilon_{r}}\frac{1}{nm} \frac{1}{r}\partial_{r}\left( r\Pi_{r\theta}\right) = 0, \label{eq:poloidal RS_sheild}
\end{equation}
where $\varepsilon_{r}$ is the neoclassical shielding factor; in Ref.~\cite{itoh1996role}, it is argued by using the fluid model that this shielding factor is $1+2q^2$, with $q$ being the safety factor.

In the gyrokinetic theory~\cite{wang2017zonal,zhang2020theory}, turbulent transport fluxes, such as the turbulent poloidal RS, the turbulent toroidal RS, and the turbulent energy flux, can be obtained by taking the corresponding velocity moment of the phase-space flux, which is calculated as $\left< \delta\dot{\bm{x}}\delta f \right>_{\text{en}}$, with $\delta \dot{\bm{x}}$ and $\delta f$ being the fluctuating particle velocity and the fluctuating distribution function.
Note that the turbulent transport fluxes are contained in the nonlinear driving term for the mesoscale ZFs in terms of the wave-wave interaction~\cite{wang2022nonlinear}.
The gyrokinetic theory~\cite{wang2017zonal,zhang2020theory} for the nonlinear driving of ZFs in a toroidal collisionless plasma indicates that 
\begin{equation} \label{eq:drive_with_sheild}
\begin{split}
    \varepsilon_{r} \partial_{t} \delta E_{r} =  &\frac{B_{\text{T}}}{nm}\frac{1}{r}\partial_{r}\left( r\Pi_{r\theta} \right) - \varepsilon_{r} \frac{B_{\text{P}}}{nm}\frac{1}{r}\partial_{r}\left( r\Pi_{r\zeta} \right) \\
    &-\varepsilon_{r} \frac{1}{ne}\partial_{r}\left[ \frac{1}{r}\partial_{r}\left( r\frac{2}{3}Q_{r} \right) \right], 
\end{split}
\end{equation}
with $\varepsilon_{r} = 1+1.64q^{2}/\sqrt{\epsilon}$ given by Refs.~\cite{rosenbluth1998poloidal,hinton1999dynamics}.
$\epsilon=r/R$, with $R$ being the major radius.
Equation~\eqref{eq:drive_with_sheild} can also be simply obtained by substituting equations~(\ref{eq:toroidal RS}-\ref{eq:turbulent energy flux}) and equation~\eqref{eq:poloidal RS_sheild} to equation~\eqref{eq:radial_force_balance}, which shows that the low frequency ZFs (LFZFs) is driven by the turbulent poloidal RS, the turbulent toroidal RS, and the turbulent energy flux.
The results obtained by the gyrokinetic theory for the cylindrical geometry agree with Eq.~\eqref{eq:drive_without_sheild}~\cite{zhang2020theory}.

Note that the factor $\varepsilon_{r}$ denotes the polarization drift enhanced by the neoclassical effect in the toroidal geometry.
If $\varepsilon_{r} = 1$ is used, it means that one ignores the neoclassical polarization effect, but retains the classical polarization effect~\cite{rosenbluth1998poloidal}; in that case, equation~\eqref{eq:drive_with_sheild} is reduced to equation~\eqref{eq:drive_without_sheild}. 
Equation~\eqref{eq:drive_with_sheild} shows that the nonlinear driving of ZFs due to the turbulent poloidal RS is shielded by the neoclassical polarization effect in the toroidal geometry, which is qualitatively consistent with the previous fluid model~\cite{itoh1996role}; however, the nonlinear driving due to the turbulent energy flux and the turbulent toroidal RS is not shielded by the neoclassical effects.

Nevertheless, one needs to investigate whether the turbulent poloidal RS would be shielded by toroidal effects and to determine the relative importance of different driving sources in the process of ZFs self-generation.
Therefore, it is of significant interest to develop a mesoscale unified picture of the nonlinear excitation of ZFs in the toroidal geometry; note that the mesoscale picture, such as the nonlinear driving due to the turbulent poloidal RS, which is widely adopted in analyzing the experiments.

In this paper, by comparing the nonlinear gyrokinetic simulation results with the above theories of ZFs, we propose a unified picture of the nonlinear driving mechanism of ZFs. We found that the low-frequency ZFs in the toroidal geometry are driven by the turbulent energy flux, the turbulent toroidal RS and the turbulent poloidal RS.
The nonlinear driving effects of the turbulent toroidal RS and the turbulent energy flux are not shielded by the toroidal effects, but the effects of the turbulent poloidal RS in the toroidal geometry is timescale dependent.
On the time scale shorter than the ion bounce time, the turbulent poloidal RS is not shielded by the neoclassical polarization factor, while on the time scale longer than the ion bounce time, the turbulent poloidal RS is shielded.

The remaining part of this paper is organized as follows: in Sec. II, we present the global nonlinear gyrokinetic simulation results, including the understanding for the turbulent poloidal RS and the turbulent energy flux; in Sec. III, we discuss the related experiments; in Sec. IV, we summarise the main findings.

\section{Nonlinear excitation of ZFs in the ITG turbulence}

\subsection{Global nonlinear gyrokinetic simulation}
To proceed, we carry out a nonlinear gyrokinetic simulation of the ITG turbulence with adiabatic electrons and kinetic ions that satisfies the nonlinear gyrokinetic Vlasov equation~\cite{brizard2007foundations},
\begin{equation} \label{vlasov}
    \partial_{t}\delta f + \{ \delta f, H_0+\delta H \}+\{f_0, \delta H\} =0.
\end{equation}
Here the subscript $0$ and the symbol $\delta$ denote the equilibrium and perturbed quantities, respectively;
$\{\cdot,\cdot\}$ is the Poisson bracket determined by equilibrium fields;
$f(r,\alpha,\theta,v_{\|},\mu,t)$ and $H$ denote the distribution function and Hamiltonian of the ion gyro-center, respectively, adopting the filed line coordinate $(r,\alpha,\theta)$. Here, $\theta$ is the poloidal angle; $\alpha = q\theta-\zeta$, with $\zeta$ being the toroidal angle.
$H_{0}=mv_{\|}^{2}/2 + \mu B + e\left<\Phi_{0}\right>_{g}$ and $\delta H = e\left<\delta\Phi\right>_{g}$, with $v_{\|}$ being the parallel velocity, $\mu$ being the magnetic moment, $B$ being the magnetic field, $\Phi$ being the electrostatic potential; $\left<\cdot\right>_{g}$ denotes the gyro-average operator.

\begin{figure}
  \centering
  \includegraphics[width=8.8cm]{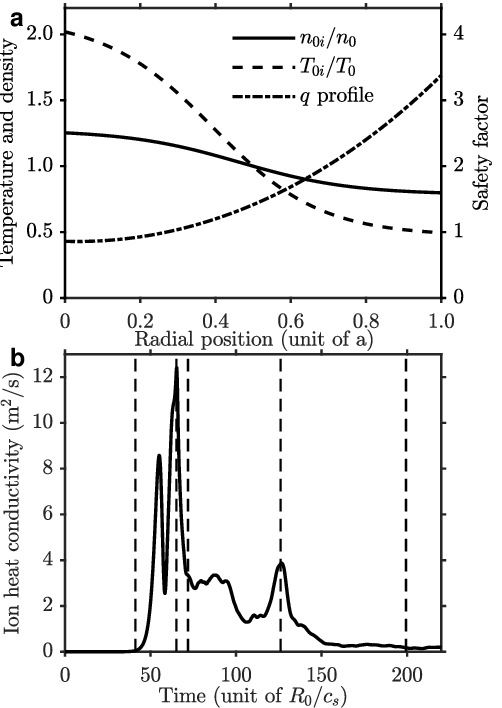}
  \caption{~\textbf{Initial equilibrium profiles and evolution.} (\textbf{a}): ion density and temperature profiles are referenced to left-hand axis; safety factor profile is referenced to right-hand axis. (\textbf{b}): ion heat conductivity at $r=0.44a$; The five vertical lines correspond to evolutionary times of $40,65,70,127,200 R_{0}/c_{s}$,respectively.}
  \label{Fig1}
\end{figure}

The global nonlinear gyrokinetic code, NLT~\cite{ye2016gyrokinetic,xu2017nonlinear,wu2024discretization}, used in this paper is based on the numerical Lie-transform method~\cite{wang2012transport,wang2013nonlinear}, which solves the nonlinear gyrokinetic equation shown in equation~\eqref{vlasov} and gyrokinetic quasi-neutrality equation~\cite{brizard2007foundations} in steps to obtain the gyro-center distribution function ($f$) and electrostatic potential ($\Phi$), as a way to advance the nonlinear evolution of the system of turbulence and LFZFs.
The numerical Lie-transform method obtains the phase space evolution of the real system by solving for the particles orbit along the unperturbed trajectory and the deviation of the real orbit from the unperturbed orbit is given by the generating vector of the Lie-transform~\cite{wang2012transport,wang2013nonlinear}.

A DIII-D-like tokamak deuterium plasma was chosen to simulate here.
The major/minor radius of the torus $R_{0}/a=1.67 \mathrm{m}/0.60 \mathrm{m}$ and the magnetic field at the magnetic axis $B_{0}=1.9 \mathrm{T}$.
The initial equilibrium profiles shown in Fig.~\ref{Fig1} are similar to the widely discussed Cyclone Base Case~\cite{dimits2000comparisons} and the key parameters are $a/\rho_{i} = 179, n_{0} = 10^{19} \mathrm{m}^{-3}, T_{0}=1.97 \mathrm{keV}$.
Here, $\rho_{i}=\sqrt{2}mc_{s}/eB_{0}$ with $c_{s}=\sqrt{T_{0}/m}$ the ion sound speed.
The simulation domain is $r \in [0, 1.0a], \alpha \in [0, 2\uppi), \theta \in [-\uppi, \uppi), v_{\|} \in [-6c_{s}, 6c_{s}], \mu B_{0}/T_{0} \in [0, 25.4]$, and the grid numbers are $N_{r} \times N_{\alpha} \times N_{\theta} \times N_{v_{\|}} \times N_{\mu} = 222 \times 190 \times 16 \times 96 \times 16$ with Gaussian discretization in $\mu$ grid while uniform discretization in other directions.
The push time step of simulation is $1/40 R_{0}/c_{s}$.
Collisional effects are not considered here because the characteristic time of ion-ion collision in our simulations ($\approx 800R_{0}/c_{s}$) is much longer than the relaxation time of ITG turbulence ($\approx 200R_{0}/c_{s}$).

In this paper, turbulent energy flux and turbulent RS are obtained by diagnosing the velocity moments of each order of distribution function, specifically,
\begin{subequations} \label{eq:diag_data}
    \begin{align}
    Q_{r} =& \left<\int {\rm d}^{3}\bm{v} \tilde{V}_{r}\delta f \cdot K \right>_{s}, \\
    \Pi_{r\zeta} =& \left<\int {\rm d}^{3}\bm{v} \tilde{V}_{r}\delta f \cdot mv_{\|}\frac{B_{\text{T}}}{B} \right>_{s}, \\
    \Pi_{r\theta} =& \left<\int {\rm d}^{3}\bm{v} f_{0}\cdot m\tilde{V}_{r}\tilde{V}_{\theta} \right>_{s},
    \end{align}
\end{subequations}
where the particles kinetic energy $K=\mu B+mv_{\|}^{2}/2$; $\left<\cdot\right>_{s}$ denotes the magnetic surface average.

\begin{figure}
  \centering
  \includegraphics[width=8.8cm]{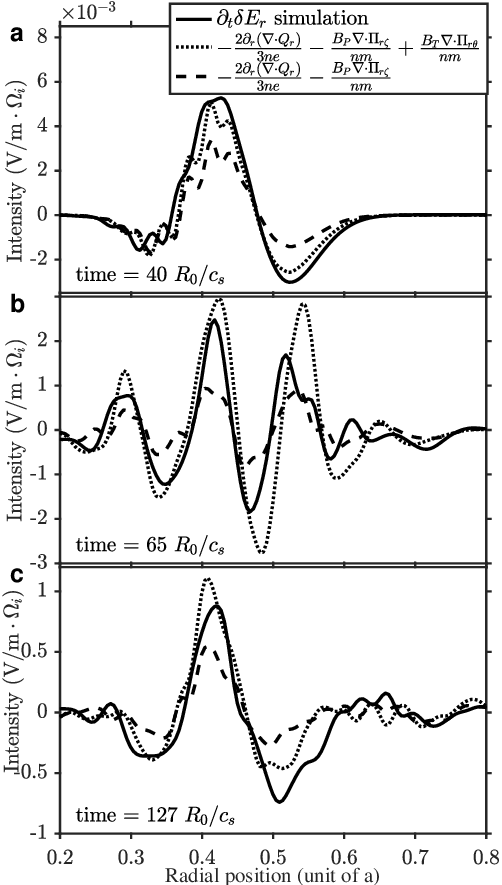}
  \caption{\textbf{Time rate of change of zonal radial electric field, $\partial_{t}\delta E_{r}$, found by the NLT nonlinear simulation, compared with theory (equation~\eqref{eq:drive_without_sheild}).}
 (\textbf{a}) The quasilinear stage, (\textbf{b}) the pre-saturation stage, and (\textbf{c}) the nonlinear turbulent burst stage, respectively, labelled on the vertical dashed lines in Fig.~\ref{Fig1}(b). The ion gyro-frequency $\Omega_{i} \approx 9.10\times 10^{7}\mathrm{s}^{-1}$.}
  \label{Fig2}
\end{figure}

Fig.~\ref{Fig2}(a) illustrates the radial structure of the zonal radial electric field and its driving sources at the quasilinear stage of ITG turbulence.
Their radial structure are almost identical and the turbulent energy flux, the turbulent poloidal RS, and the turbulent toroidal RS contribute approximately $50\%, 35\%$ and $5\%$ to the zonal radial electric field, respectively, which agrees well with equation~\eqref{eq:drive_without_sheild}, hence both the turbulent poloidal RS and the turbulent energy flux are important in nonlinearly driving ZFs in quasilinear stage.
Further, Fig.~\ref{Fig2}(b-c) shows the radial structures during the nonlinear saturation and the nonlinear burst stage of turbulence.
When the turbulent poloidal RS is not considered in the driving sources, the radial structure is far from the time-varying rate of zonal radial electric field, whereas the two are obviously more in line with each other once the contribution of turbulent poloidal RS is taken into account.
Thus it is obvious that the turbulent poloidal RS is important for the nonlinear driving of ZFs in these three stages.
The nonlinear driving effects of the turbulent toroidal RS is relatively small compared to that of the turbulent poloidal RS or the turbulent energy flux, and is therefore not discussed subsequently.

Note that the diagnosis data here has been filtered by using the time-domain low-pass filtering to eliminate oscillations above the ion bounce frequency.
This operation is motivated by two considerations: firstly, the driving equation for ZFs is only applicable to the low-frequency situation, due to the orbit average is used in the derivation of gyrokinetic theory~\cite{rosenbluth1998poloidal,wang2017zonal};
secondly, based on the fundamental theory~\cite{biglari1990influence,hahm1999shearing} that the $\bm{E}\times\bm{B}$ shear flows distort the turbulence eddies and reduce their correlation, the effects of high-frequency oscillations are much weaker than the LFZFs.

Fig.~\ref{Fig2} seems to imply that the effect of the turbulent poloidal RS on driving ZFs is not shielded by the toroidal effects, which is similar to the behaviour in the cylindrical geometry (equation~\eqref{eq:drive_without_sheild}).

\subsection{The neoclassical shielding factor of the turbulent poloidal RS}
The nonlinear driving equation of LFZFs shown in equation~\eqref{eq:drive_without_sheild} is well confirmed during these three stages shown in Fig.~\ref{Fig2} (quasilinear growth of turbulence, nonlinear initial saturation, and nonlinear bursting), which indicates no neoclassical shielding to the turbulent poloidal RS.
Further diagnostic data from ITG turbulence, however, suggest that in the case of turbulence quenching or quasi-steady-state, neither equation~\eqref{eq:drive_without_sheild} nor~\eqref{eq:drive_with_sheild} is satisfied, and it even appears that the phases of $\nabla \cdot \Pi_{r\theta}$ and $\partial_{t}\delta E_{r}$ are opposite. 
We integrate equation~\eqref{eq:drive_without_sheild} to find
\begin{equation} \label{eq:Er_without_sheild}
    \delta E_{r}(t) = \frac{1}{ne}\partial_{r}\delta p_{i} + \delta u_{\zeta}B_{\text{P}} + \int_{-\infty}^{t} \frac{B_{\text{T}}}{nm}\frac{1}{r}\partial_{r} \left( r\Pi_{r\theta} \right) {\rm d} t'.
\end{equation}
It should be point out that $\delta p_{i}$ and $\delta u_{\zeta}$ in equation~\eqref{eq:Er_without_sheild} are calculated in kinetic form, that is, using the second-order energy moment and the first-order parallel velocity moment of the perturbation distribution function, respectively.
For simplicity of notation, the last term on the right-hand side of equation~\eqref{eq:Er_without_sheild} is formally noted as $-\delta u_{\theta} B_{\text{T}}$.

\begin{figure}
  \centering
  \includegraphics[width=8.8cm]{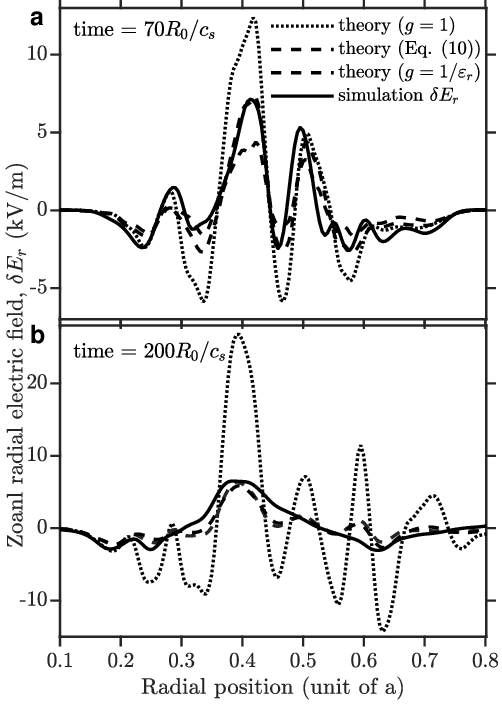}
  \caption{\textbf{Zonal radial electric field, $\delta E_{r}$, found by the NLT nonlinear simulation, compared with the theoretical model  (equations~(\ref{eq:poloidal RS_response}-\ref{eq:example_responce})). } \textbf{(a)}: $t=70 R_{0}/c_{s}$ (when turbulence is in the decreasing stage after saturation) and \textbf{(b)}: $t=200 R_{0}/c_{s}$ (when turbulence has quenched). Here, the parameters in equation~\eqref{eq:example_responce} are set to $y=0.5$, and $\tau_{b} \approx 2\uppi q/\sqrt{\varepsilon_{r}}\cdot R_{0}/c_{s}$ with $\varepsilon_{r} \approx 6 $.}
  \label{Fig3}
\end{figure}

It is confusing that the fluid poloidal flow ($\approx 25 \mathrm{kV/m}$) nonlinearly driven by the time-integrated turbulent poloidal RS is significantly larger than the turbulence-driven zonal radial electric field ($\approx 7 \mathrm{kV/m}$) throughout the whole relaxation evolution, as shown by the dotted line in Fig.~\ref{Fig3}(b);
and after considering the neoclassical shielding factor in the toroidal geometry (the last term on the right-hand side of equation~\eqref{eq:Er_without_sheild} divided by $\varepsilon_{r}$), the nonlinear driving equations for LFZFs shown in equation~\eqref{eq:drive_with_sheild} likewise fit well again, as is shown by the dashed line in Fig.~\ref{Fig3}.
To have a unified picture, we modify the last term in equation~\eqref{eq:Er_without_sheild}, which is essentially $-\left<\delta u_{\theta}\right> B_{\text{T}}$, and equation~\eqref{eq:Er_without_sheild} is modified to
\begin{equation} \label{eq:poloidal RS_response}
\delta E_{r}(t) = \frac{1}{ne}\partial_{r}\delta p_{i} + \delta u_{\zeta}B_{\text{P}} + B_{\text{T}}\int_{-\infty}^{t} \frac{1}{nm}\frac{1}{r}\partial_{r}\left[ r\Pi_{r\theta}(t') \right] \cdot g(t-t') {\rm d}t'.
\end{equation}
Here, the usual definition, $3\delta p_{i}/2 = \int {\rm d}^{3}\bm{v}\left<\delta f\right>_{s} K$, is used to compute the ion pressure perturbation from the simulation data.
The response function $g(\tau)$ in equation~\eqref{eq:poloidal RS_response} should satisfy:
\begin{equation*} \label{eq:response_function}
    g(\tau) = \begin{cases}
        1, & \text{if } \tau/\tau_{b} \to 0; \\
        1/\varepsilon_{r}, & \text{if } \tau/\tau_{b} \to \infty.
    \end{cases}
\end{equation*}
In this paper, we propose the following response function:
\begin{equation} \label{eq:example_responce}
    g(\tau) = \frac{1}{\varepsilon_{r}} + \left( 1-\frac{1}{\varepsilon_{r}} \right){\rm e}^{-\frac{\tau}{y \tau_{b}}},
\end{equation}
with $\tau_{b}$ being the bounce time for trapped particles, and $y$ being a numerical factor, which is chosen as $y=0.5$ here.

Following the above discussions, we unify equations~\eqref{eq:drive_without_sheild} and ~\eqref{eq:drive_with_sheild} into 
\begin{equation} \label{eq:drive_sheild}
\begin{split}
    \partial_{t} \delta E_{r} =&~-\frac{1}{ne}\partial_{r}\left[ \frac{1}{r}\partial_{r}\left( r\frac{2}{3}Q_{r} \right) \right]  - \frac{B_{\text{P}}}{nm}\frac{1}{r}\partial_{r}\left( r\Pi_{r\zeta} \right)  \\
    &+\partial_{t}\int_{-\infty}^{t} \frac{B_{\text{T}}}{nm}\frac{1}{r}\partial_{r}\left[ r\Pi_{r\theta}(t') \right] \cdot g(t-t') {\rm d}t,
\end{split}
\end{equation}
with $g(\tau)$ given by equation~\eqref{eq:example_responce}. Note that in the limit $g(\tau) \to 1$, equation~\eqref{eq:drive_sheild} is reduced to equation~\eqref{eq:drive_without_sheild}; in the limit $g(\tau) \to 1/\varepsilon_{r}$, equation~\eqref{eq:drive_sheild} is reduced to equation~\eqref{eq:drive_with_sheild}.

The NLT simulation results of the zonal radial electric field is compared to the unified theoretical model in Fig.~\ref{Fig3}.
It can be seen that the simulation results agree well with the theoretical model given by equations~(\ref{eq:poloidal RS_response}-\ref{eq:example_responce}). 
However, Fig.~\ref{Fig3}(a) shows that in the post-saturation stage, neither the model with $g(\tau) = 1$ (corresponding to equation~\eqref{eq:drive_without_sheild}) nor the model with
$g(\tau) = 1/\varepsilon_{r}$ (corresponding to equation~\eqref{eq:drive_with_sheild}) agrees with the simulation results.
Fig.~\ref{Fig3}(b) shows that in the turbulence quench stage, the model with $g = 1/\varepsilon_{r}$ agrees well with the simulation results, while the model with $g = 1$ does not.

The response function in equation~\eqref{eq:example_responce} represents the shielding of the turbulent poloidal RS in the toroidal geometry.
In the limit $\tau \gg \tau_{b}$, equation~\eqref{eq:drive_sheild} is reduced to equation~\eqref{eq:drive_with_sheild}; this means that on the time scale longer than the ion bounce time, the neoclassical polarization effect emerges and the nonlinear driving of the poloidal RS is shielded; this corresponds to the turbulence quench stage, which is shown in Fig.~\ref{Fig3}(b).
In the limit $\tau \ll \tau_{b}$ but larger than the ion gyro-period, equation~\eqref{eq:drive_sheild} is reduced to equation~\eqref{eq:drive_without_sheild}; this means that on this short time scale, the neoclassical polarization cannot come into play intime, and the nonlinear driving of the poloidal RS is not shielded; this corresponds to the rapid growth stages, which is shown in Fig.~\ref{Fig2}.

\subsection{The orbital average in calculating the turbulent energy flux on the long time scale}
In this subsection, we demonstrate that the orbital average is needed in calculating the ion pressure from the turbulent energy flux by using the simulation data on the long time scale.

The dispersion relation of LFZFs shown in equation~\eqref{eq:drive_with_sheild} relates the turbulent energy flux to the zonal radial electric field.
The ion pressure gradient in equation~\eqref{eq:Er_without_sheild} can be related to the turbulent energy flux by using equation~\eqref{eq:turbulent energy flux}.
However, it should be pointed out that the phase-space turbulent energy flux should be orbital averaged when examining the mesoscale dynamics of ZFs in gyrokinetic theory~\cite{wang2017zonal}; therefore, the orbital average should be done 
when the ion pressure is calculated by using the turbulent energy flux.
Following Ref.~\cite{wang2017zonal}, the nonlinear driving of ZFs on the long time scale due to the turbulent energy flux is through the ion pressure term given by
\begin{equation} \label{eq:bounce_turbulent energy flux}
    \left<\delta p_{i}\right> = -\int {\rm d}^{3}\bm{v} \left< (\frac{K}{T_{i}}-\frac{3}{2}) \frac{2f_{0}}{3nT_{i}}\int {\rm d}t \frac{1}{r}\partial_{r} (r\frac{2}{3}Q_{r}) \right>_{orb.} K.
\end{equation}
Here, $\left< \cdot \right>_{orb.}$ denotes the orbital average~\cite{wang2017zonal,wang2024self}.

\begin{figure}
  \centering
  \includegraphics[width=8.8cm]{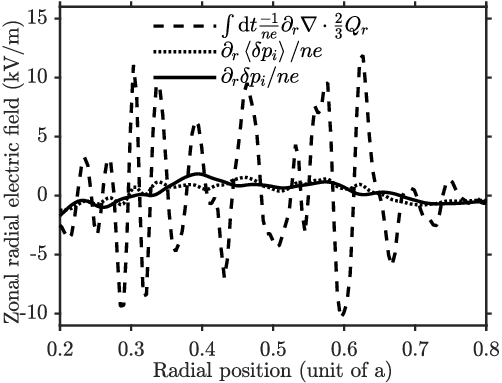}
  \caption{\textbf{The ion pressure gradient and the turbulent energy flux.} The solid line denotes the results found by using the usual definition of ion pressure; the dotted line represents the results found by using equation~\eqref{eq:bounce_turbulent energy flux}, and the time integration interval in equation~\eqref{eq:bounce_turbulent energy flux} was selected for the entire turbulent relaxation process ($0 - 200 R_{0}/c_{s}$).}
  \label{Fig4}
\end{figure}

As shown in Fig.~\ref{Fig4}, 
the ion pressure calculated by using equation~\eqref{eq:bounce_turbulent energy flux} agrees well with the usual definition of ion pressure; while the ion pressure calculated by using equation~\eqref{eq:bounce_turbulent energy flux} with the orbital average operator ignored disagrees with the usual definition.

\section{Discussion on the related experiments}
\subsection{JFT-2M Tokamak}
In the LCO before L-H transition of JFT-2M, one of the reasons why ZFs do not exist during LCO is that the turbulent poloidal RS is found to be too small to explain the oscillatory flow, by considering the dielectric constant $\varepsilon_{r}$.
In these evaluations~\cite{kobayashi2013spatiotemporal}, the turbulent poloidal RS per unit mass density, measured on the time scale $\tau_{LCO}$, with $\tau_{LCO}$ the period of LCO, was $|\Pi_{r\theta}| \approx 0.7\times 10^{5} \mathrm{m^{2}/s^{2}}$ and the modulation of $\bm{E}\times\bm{B}$ velocity in the LCO was $|\delta u_{\theta} | \approx 500 \mathrm{m/s}$.
Ref.~\cite{kobayashi2013spatiotemporal} explains that after taking into account $\varepsilon_{r} \approx 20 $ in the plasma edge, the amplitude of modulation poloidal flow driven out by the effects of the turbulent poloidal RS is estimated to be $15 \mathrm{m/s}$, which is much smaller than $|\delta u_{\theta}|$.

However, it is important to point out that the measured turbulent poloidal RS on the time scale shorter than $\tau_{LCO}$ is not yet shielded by toroidal effects.
Using the parameters given in Ref.~\cite{kobayashi2013spatiotemporal}, one finds $\tau_{LCO} \approx 0.22~\mathrm{ms}$ and $\tau_{b} \approx 2\pi qR/(\sqrt{\epsilon}v_{th,i}) \approx 0.46~\mathrm{ms}$.
Therefore the neoclassical shielding factor does not need to be considered, which corresponds to $g(\tau) \approx 1$ in equation~\eqref{eq:poloidal RS_response}.
Thus, without considering the shielding factor $\varepsilon_{r}$, the poloidal flow velocity nonlinearly driven by the turbulent poloidal RS should be approximately $300 \mathrm{m/s}$ instead of $15 \mathrm{m/s}$.
This velocity is approximately $60\%$ of the measured poloidal velocity $|\delta u_{\theta}|$ balanced with the zonal radial electric field, and further this ratio is similar to our simulation results, where the turbulent poloidal RS contributes about $35\%$ in the quasilinear stage without considering shielding effects (Fig.~\ref{Fig2}), and about $65\%$ throughout the relaxation process with considering shielding effects (Fig.~\ref{Fig3}).
These discussions can be used to explain why the turbulent poloidal RS appears to be so small on JFT-2M.

\subsection{HL-2A Tokamak} 
In the Low-Intermediate-High (L-I-H) confinement transitions on HL-2A Tokamak, two types of LCOs are found, namely type-Y LCO and type-J LCO~\cite{cheng2013dynamics}.
The former appears first after an L-I transition and the turbulent poloidal RS plays an essential role in the Type-Y, which is consistent with the predator-prey model~\cite{kim2003zonal}. In contrast, type-J with the opposite plausible causality between the turbulence and zonal radial electric field is dominated by ion-pressure-gradient-induced drift.
More importantly, an I-H transition is demonstrated to occur only from type-J.

These findings from HL-2A Tokamak imply that the importance of turbulent Reynolds stress for nonlinear driving of LFZFs varies across time.
Note that during the L-I-H transitions, there are strong turbulent fluctuations of floating potentials in the L-mode, relatively weak in the I-phases, and the weakest turbulence in the H-mode, as shown in Fig.2 of Ref.~\cite{cheng2013dynamics}.
Thus in the L-I transition, the strong turbulent outburst corresponds to the first limit in equation~\eqref{eq:poloidal RS_response}, i.e., the fluid poloidal flow nonlinearly driven by the turbulent poloidal RS is not shielded by the neoclassical polarization effects, and so it is natural to observe that the type-Y LCO is dominated by the turbulent poloidal RS at L-I transition. 
In contrast, when the turbulent fluctuations are weak, there is enough time for the turbulent poloidal RS to be weakened to a residual level, and thus no correlation between the turbulent poloidal RS and the zonal radial electric field is observed in type-J LCO.
Considering that the turbulent poloidal RS involves an ensemble average operator after the product of perturbed radial wave number and poloidal wave number, it is not difficult to understand that no significant turbulent poloidal RS can be measured during stages of incoherent steady-state turbulence.
However, turbulent energy flux involves the poloidal wave number and the perturbed ion pressure, both of which  still maintain a fixed phase difference during the steady-state turbulence stage, which can be analogized to a typical diffusion process, allowing for the observation that type-J is dominated by the turbulent energy flux.

\section{Summary}
In summary, by comparing the nonlinear gyrokinetic simulation results with the previous theories of ZFs, we proposed a unified mesoscale picture of the nonlinear driving mechanism of ZFs in the toroidal geometry.

The zonal radial electric field is nonlinearly driven by the turbulent energy flux, toroidal RS, and poloidal RS, in the ITG turbulence.
The nonlinear driving effects due to the turbulent energy flux and toroidal RS are not shielded by the neoclassical polarization effect.
However, the nonlinear driving effect of the turbulent poloidal RS in the toroidal geometry is time-scale dependent.
When the time scale considered is shorter than the ion bounce period, the effect of the turbulent poloidal RS is not shielded by the toroidal effects, and its behaviour is the same as in the cylindrical geometry.
When the time scale considered is longer than the ion bounce period, the effect of the turbulent poloidal RS is shielded by the neoclassical polarization effects.
This is due to the following fact.
On the time scale longer than the ion gyro-period, the classical polarization effect comes into play;
on the time scale longer than the ion bounce period, the neoclassical polarization effect comes into play.
To describe the dynamics of ZFs involving the two time scales in the toroidal geometry, we proposed equations~(\ref{eq:poloidal RS_response}-\ref{eq:drive_sheild}).

We point out that the effect of the turbulent poloidal RS measured in the JFT-2M device~\cite{kobayashi2013spatiotemporal} should not have been shielded, due to the fact that the time scale of the LCO observed there is shorter than the ion bounce period.
The two types of the LCO in the L-I-H transition observed in the HL-2A device~\cite{cheng2013dynamics}, i.e., type-Y with strong turbulent fluctuations and type-J with rather weak turbulence, correspond to the two limits of the turbulent poloidal RS driving effects, respectively.
In the type-Y LCO, the turbulent poloidal RS is not shielded by the toroidal effects; in contrast, in the type-J LCO, the driving effects of the turbulent poloidal RS is shielded, while the ion-pressure-gradient nonlinearly driven by the turbulent energy flux becomes important.
These findings are important for understanding the LCO experiments near the L-H transition threshold and resolve the controversial issues that are fundamental.
Finally, we point out that in a steady-state turbulence, the turbulent poloidal RS may not be important in nonlinear driving ZFs, since it is shielded by the neoclassical effects in the toroidal geometry on the time scale longer than the ion bounce period.

\section*{Declarations}
\begin{itemize}
\item Funding: This work was supported by the National MCF Energy R\&D Program of China under Grant No. 2019YFE03060000, and the National Natural Science Foundation of China under Grant No. 12075240.
\item Competing interests: The authors declare no competing interests. 
\item Code availability: The data that support the findings of this study are available from the corresponding author upon reasonable request.
\end{itemize}

\section*{References}

\bibliography{ref}

\clearpage
\end{document}